\documentclass[onecolumn,sort&compress,numbers]{els-mrw} 
\usepackage{amsmath,amssymb,amsfonts,amsthm,makeidx,graphicx}
\usepackage{dsfont,bm}
\usepackage[greek,english]{babel}
\usepackage{textgreek}
\usepackage{txfonts}
\usepackage{helvet}
\usepackage[colorlinks,citecolor=blue,linktoc=all,linkcolor=cyan]{hyperref}
\begin{document}
%%Please add any additional required packages before this commented line.
%%%%%%%%%%%%%%%%%%%%%%%%%%%%%%%%%%%%%%%%%%%%%%%%%%%%%%%%%%%%%%%%
%% the following items are mandatory: 
%% - title
%% - author names
%% - affiliation details
%% - abstract
%% - keywords
%% Precise, concise, and informative description of the focus of this work. Avoid abbreviations and formulae in the title
\chapter{Theory of resonances}\label{chap1}
\author[1,2]{Maxim Mai}%
\address[1]{
	\orgname{University of Bern}, 
	\orgdiv{Albert Einstein Center for Fundamental Physics, Institute for Theoretical Physics},
	\orgaddress{Sidlerstrasse 5, Bern, 3012, Switzerland}}
\address[2]{
	\orgname{The George Washington University},
	\orgdiv{Department of Physics},
	\orgaddress{725 21st St, NW 20052, District of Columbia, USA}}
\articletag{Account of excited hadrons with a focus on history, theoretical tools and new developments that serves pedagogical or review purposes.}
%%%%%%%%%%%%%%%%%%%%%%%%%%%%%%%%%%%%%%%%%%%%%%%%%%%%%%%%%%%%%%%%
\maketitle
%%%%%%%%%%%%%%%%%%%%%%%%%%%%%%%%%%%%%%%%%%%%%%%%%%%%%%%%%%%%%%%%

%%%%%%%%%%%%%%%%%%%%%%%%%%%%%%%%%%%%%%%%%%%%%%%%%%%%%%%%%%%%%%%%
%% the following item is mandatory: 
%% 100-150 word summary of the chapter
\begin{abstract}[Abstract]
	We give a pedagogical introduction to the theory of resonances, focusing specifically on the spectrum of excited strongly interacting particles. After providing historical context starting from the atomic theory, we summarize the status of theoretical and experimental research. We discuss then, the methodology to determine universal resonance parameter through the analytical properties of the transition amplitudes. For this, the main aspects of the $S$-matrix theory are introduced including some explicit calculations in simplified cases. In the last section, we summarize several frequently used amplitude types also making a connection to Effective Field Theories, as well as provide first glimpse into some more advanced applications to further resonance properties.
\end{abstract}
%%%%%%%%%%%%%%%%%%%%%%%%%%%%%%%%%%%%%%%%%%%%%%%%%%%%%%%%%%%%%%%%

%%%%%%%%%%%%%%%%%%%%%%%%%%%%%%%%%%%%%%%%%%%%%%%%%%%%%%%%%%%%%%%%
%% 5-10 words that embody the key topics in the chapter. What terms would someone put into a search engine if they were looking for a chapter like this?
\begin{keywords}
	Hadron spectrum\sep 
	amplitudes\sep 
	excited states \sep 
	scattering \sep 
	Breit-Wigner \sep
	Flatt{\'e} \sep
	resonances\sep
	Transition form factors \sep
	unitarity \sep
	S-matrix
\end{keywords}
%%%%%%%%%%%%%%%%%%%%%%%%%%%%%%%%%%%%%%%%%%%%%%%%%%%%%%%%%%%%%%%%

%%%%%%%%%%%%%%%%%%%%%%%%%%%%%%%%%%%%%%%%%%%%%%%%%%%%%%%%%%%%%%%%
\section*{Objectives}
\begin{itemize}
	\item Historical context: search for the building blocks of nature and discovery of excited states of matter.
	\item Status of the experimental and theoretical research unraveling spectrum of strongly interacting particles. 
	\item Concept of the universal parameters of resonances and unification thereof with bound states through $S$-matrix formalism. 
	\item Introduction into mathematical principles of the $S$-matrix theory.
	\item Breit-Wigner, $K$-matrix, Flatt{\'e}, Chiral unitary parametrizations of the scattering amplitudes.
	\item Compositeness of states, Chiral trajectories, transition form factors of resonances.
\end{itemize}
%%%%%%%%%%%%%%%%%%%%%%%%%%%%%%%%%%%%%%%%%%%%%%%%%%%%%%%%%%%%%%%%

%%%%%%%%%%%%%%%%%%%%%%%%%%%%%%%%%%%%%%%%%%%%%%%%%%%%%%%%%%%%%%%%
\section{Introduction -- Hadrons, resonances, excited states}
\label{sec:intro}
%%%%%%%%%%%%%%%%%%%%%%%%%%%%%%%%%%%%%%%%%%%%%%%%%%%%%%%%%%%%%%%%

For the longest part of humanity's history the perception of nature was based on experience with everyday objects which cover length scales of $\sim10^0~\rm m$. It is only later and thanks to the modern scientific method that we have been successively \emph{zooming} onto smaller and smaller building blocks of nature, truly searching for the indivisible (Greek \textgreek{ἄτομον}) particles building up the world around us. The word ``atom'' was introduced into modern sciences through analysis of masses of chemical elements appearing ar ratios of whole numbers~\cite{Dalton1808}. With the discovery of electron~\cite{Thomson:1897cm} and later of the atomic nucleus~\cite{Rutherford:1911zz}, atoms turned out later as not indivisible either. While the electron is in the modern understanding indeed indivisible, the atomic nucleus consists of protons and neutrons. Just as the name choice suggests (\textgreek{πρῶτον} Greek for ``first'') the proton indeed was the first representative of a large class of particles --  the so called \emph{hadrons} (Greek \textgreek{ἁδρός} for ``strong'') -- id est subatomic particles bound by and interacting through the strong force. The fundamental/elementary degrees of freedom of this force being quarks and gluons, proposed and discovered in the second half of the last century~\cite{Gell-Mann:1964ewy,Zweig:1964ruk,Bloom:1969kc}. In a first approximation, the quark model classified hadrons as mesons and baryons as bound states of quark-antiquark pair and three quarks, respectively. %For more details and context of strong interaction as one of four fundamental forces of nature according to the standard model of particle physics see Chapter~{\color{red}10003,10004}.

\begin{figure}[hbt]
	\centering
	\includegraphics[width=\linewidth]{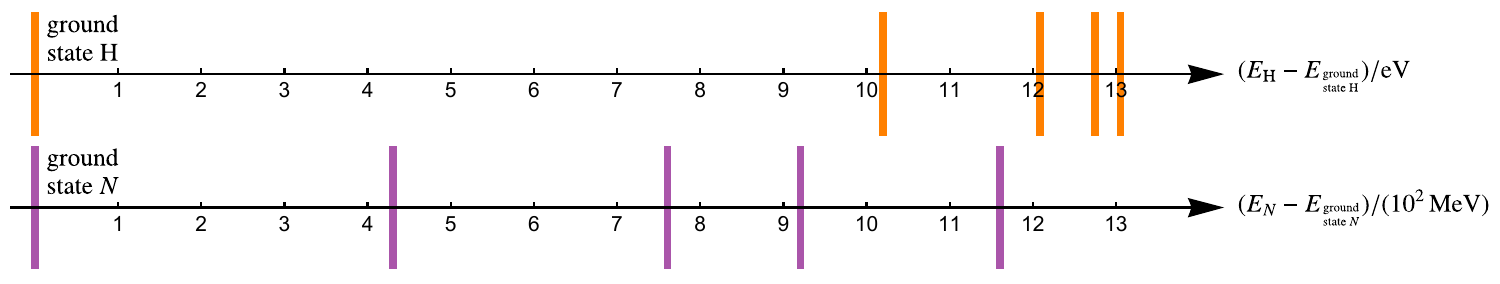}
	\caption{Relative energies of the excited states of Hydrogen atom (top) and Nucleon (bottom) with respect to the respective ground states. Hydrogen states are calculated through the Rydberg formula while the nucleon excited states ($J^{P}=1/2^+$) are taken as central values from the summary tables of the Particle Data Group~\cite{ParticleDataGroup:2024cfk}.}
	\label{fig:Hydrogen-vs-Nucleon}
\end{figure}

The search for the indivisible building blocks of matter around us outlined above has led us to an interesting fact. Already at scales of atoms (length scales $\sim10^{-10}~\rm m$) but even more so for hadrons (length scales $\sim10^{-15}~\rm m$) concepts of classical physics need to be replaced with quite non-intuitive language of quantum mechanics. For a historical reflection of early disputes on nature and philosophical implications of quantum mechanics see, e.g., W.~Heisenberg's \emph{``Physics and Beyond: Encounters and Conversations''}~\cite{HeisenbergBOOK}. As described there, the mathematical language of quantum mechanics was initially heavily disputed providing, however, the only viable explanation to a series of experiments. Among others, one of the major breakthroughs of quantum mechanics is associated with the correct description of the pattern of excited states of atomic spectra. Today calculation of the excited states of the hydrogen atom belongs to the standard curriculum of the introductory courses on quantum mechanics. As we have seen before, hadrons are also composite and, thus, it is only reasonable to assume that they also do not only exist in the ground states but build a tower of unstable states, which can be excited and decay in accordance with principles of quantum mechanics. For a side-by-side comparison of the hydrogen and proton spectra see Fig.~\ref{fig:Hydrogen-vs-Nucleon}. Currently, it is believed that connecting fundamental theory of strong interaction with experimental observations of excited states (usually referred to as \emph{resonances}~\cite{Wigner:1946zz}) of hadrons is a litmus test of our understanding of strong interaction, analogously to the past experience with the atomic spectra. This research direction is referred to as the \emph{the hadron spectroscopy} and is the main aspect of the present manuscript.

%%%%%%%%%%%%%%%%%%%%%%%%%%%%%%%%%%%%%%%%%%%%%%%%%%%%%%%%%%%%%%%%
%%%%%%%%%%%%%%%%%%%%%%%%%%%%%%%%%%%%%%%%%%%%%%%%%%%%%%%%%%%%%%%%

%%%%%%%%%%%%%%%%%%%%%%%%%%%%%%%%%%%%%%%%%%%%%%%%%%%%%%%%%%%%%%%%
\section{From theory to experiment and back}
\label{sec:theory-exp}
%%%%%%%%%%%%%%%%%%%%%%%%%%%%%%%%%%%%%%%%%%%%%%%%%%%%%%%%%%%%%%%%

Theoretically, the only known way to reconcile principles of quantum mechanics with those of special relativity is the notion of quantum field theories. For an early review of the matter see Ref.~\cite{Wigner:1957ep}, for philosophical aspects see \cite{Talagrand_2022}. Going beyond quantum mechanics, such theories allow for creation of particles and antiparticles, relevant at the energy scales of hadron physics ($\rm MeV$). Quantum field theory of strong interaction is called Quantum Chromodynamics (QCD)~\cite{Gross:1973id,Politzer:1973fx}. As such it provides many high precision predictions tested for example at the LHC or HERA, for current status see review tables of Ref.~\cite{ParticleDataGroup:2024cfk}. While QCD has passed all these tests for over half a century now, direct access to properties of excited hadrons (their masses, widths or branching ratios) is obscured by the simple fact that the QCD coupling constant is large for small energies. Thus, the well established perturbative/diagrammatic approach is doomed to fail. The reason for this lies in the very structure of QCD, being a non-abelian gauge theory. In such theories, gluons (gauge bosons of theory) self-interact which leads to the antiscreening of the interaction coupling. As a matter of fact the increase of the strong coupling at low energies gives rise to the \emph{confinement hypothesis} which states that no matter- or gauge-fields of QCD can be observed directly.%{\color{red}10011}. 

As of today, the most commonly used theoretical approaches aiming to circumvent the above mentioned challenges and providing access to hadron spectrum can be subdivided into three classes: (1) numerical, (2) functional and (3) effective approaches\footnote{The separation and naming is not canonical but is utilized here simply for introductory reasons.
% Dedicated chapters shall be consulted for more details on each of the approaches.
}. 
%%%
(1) Numerical lattice gauge theory approach (short Lattice QCD) provide currently the most direct access to the low-energy structure of QCD in a non-perturbative fashion. Such calculations are performed on a discretized Euclidean space-time, in finite volume and typically at unphysical quark masses. These technicalities do not allow one to directly compare Lattice QCD results with the phenomenological ones. However, methods exists to systematically deal with these issues,
% as described in %Chapter~{\color{red}20019}. For further review of hadron spectroscopy through Lattice QCD 
see, e.g., recent review~\cite{Mai:2022eur}. 
%%%
(2) Functional methods relate but go beyond the philosophy of the original quark models describing baryons as relativistic three-quark bound states within QCD. In many cases such approaches, indeed, provide a simultaneous description of large parts of hadron spectrum throughout large energy ranges, see 
%Chapter {\color{red}10028} and 
Ref.~\cite{Eichmann:2016yit} for a dedicated review. 
%%%
(3) The last class of approaches follows a different trajectory, relying on the scale separation between the heavy (charm, bottom, top) and light (up, down, strange) degrees of freedom. Through a well-defined methodology the heavy degrees of freedom are integrated out leaving one with an effective field theory, which encodes the dynamics relevant to the low-energy regime. In the case of QCD, the low-energy effective theory is referred to as the Chiral Perturbation Theory (CHPT) necessarily formulated through the lightest bosons in the hadron spectrum, the pseudoscalar mesons $\{\pi^0,\pi^+,\pi^-,K^0,K^-,K^+,\bar K^0,\eta\}$. Baryons can be introduced as heavy matter fields in a fully systematic way, i.e., respecting the symmetries of QCD. Since its advent around 1980's~\cite{Weinberg:1978kz,Gasser:1984gg, Gasser:1983yg} CHPT has become a powerful tool and in many cases as a benchmark in the threshold and subthreshold energy region, see \cite{Bernard:2006gx,Bernard:2007zu,Scherer:2002tk,Meissner:1993ah}.
% as well as Chapter~{\color{red}20009}. 
Regarding the hadron spectrum, CHPT cannot be applied in the original sense of perturbation theory as resonances are truly non-perturbative phenomena as we will see below. However, methods to extend the perturbative approach through the principles of $S$-matrix theory exists, for a related reviews see Refs.~\cite{Pelaez:2015qba,Mai:2022eur}.

%%%%%%%%%%%%%%%%%%%%%%%%%%%%%%%%%%%%%%%%%%%%%%%%%%%%%%%%%%%%%%%%
\begin{figure}[ht]
	\centering
	\includegraphics[width=\linewidth]{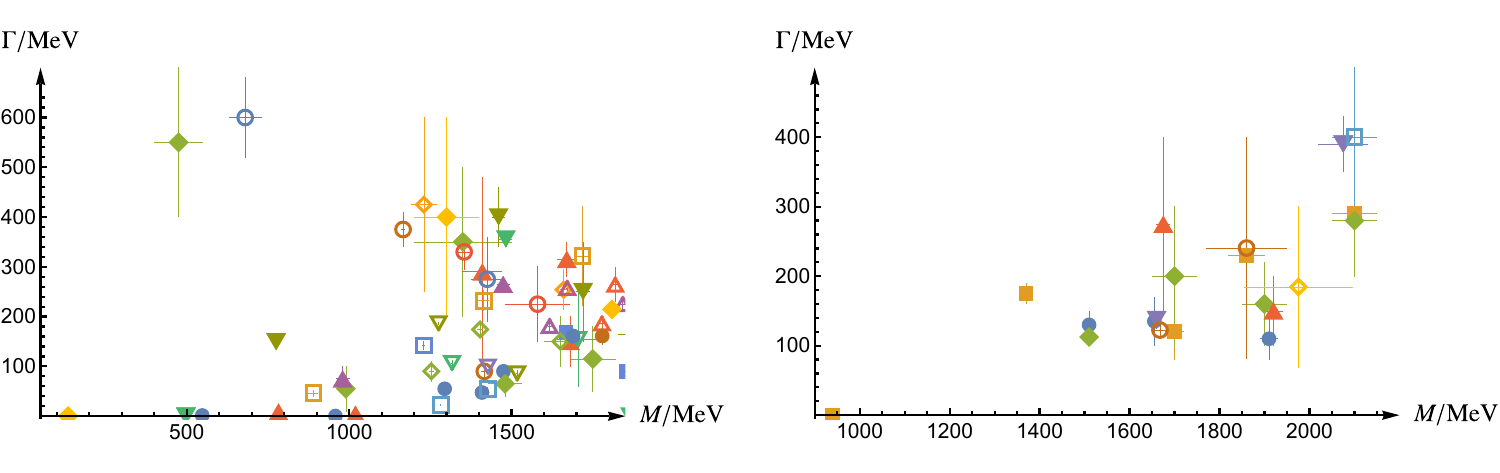}
	\caption{Sample of masses and widths of mesons and baryons~\cite{ParticleDataGroup:2024cfk}. Different symbols represent states of the same quantum numbers (total angular Momentum, Parity, charge conjugation). Vertical/horizontal bars depict the uncertainty of mass and width determination, respectively. Compare also Fig.~\ref{fig:Hydrogen-vs-Nucleon} with respect to nucleon excited states depicted here by orange squares.}
	\label{fig:MES-BAR-Resonances}
\end{figure}
%%%%%%%%%%%%%%%%%%%%%%%%%%%%%%%%%%%%%%%%%%%%%%%%%%%%%%%%%%%%%%%%

\begin{figure}[ht]
	\centering
	\includegraphics[width=0.6\linewidth]{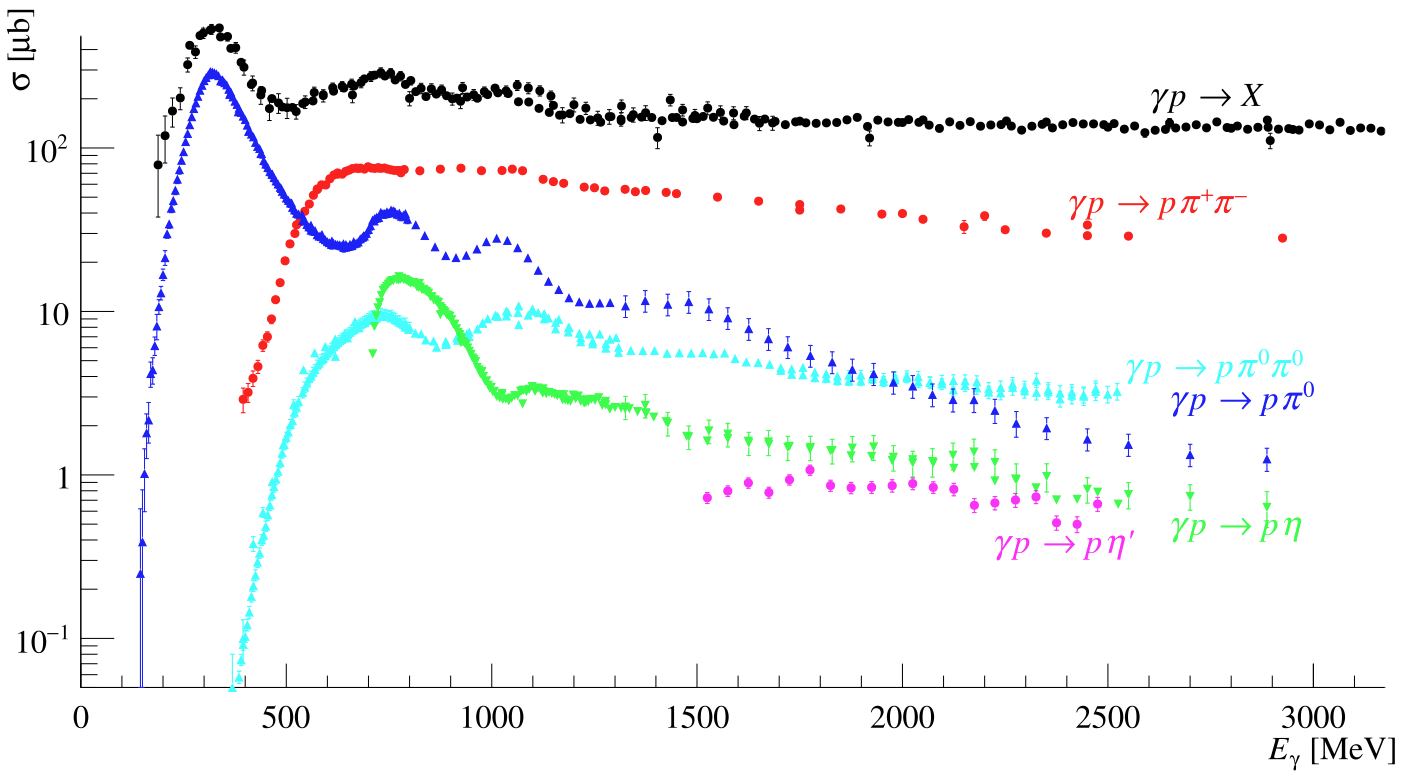}
	\caption{World (status 2024) cross section data on photoproduction experiments~\cite{Braghieri:1994rf,Aachen-Hamburg-Heidelberg-Munich:1975jed,CB-ELSA:2003rxy,CB-ELSA:2004sqg,Wu:2005wf,Thoma:2007bm,CBELSATAPS:2009ntt,Schumann:2010js,CrystalBallatMAMI:2010slt} as a function of photon energy ($E_\gamma$). Figure adapted from review~\cite{Thiel:2022xtb}.}
	\label{fig:exp-data}
\end{figure}

As of the experimental situation, our current knowledge is collected by the Particle Data Group (PDG)~\cite{ParticleDataGroup:2024cfk} which compiles all experimental and theoretical evidences for new excited states. Confirmed baryon states receive a 4 star rating, first evidences are marked by one star. As of now PDG lists around 150 excited mesons and around 50 confirmed excited baryons. Most of these states are indeed excited/unstable\footnote{In this article we disregard weak decays, thus, e.g., pions, kaons or etas are seen as stable as proton or neutron.} characterized through two real-valued numbers -- the mass ($M\in\mathds{R}$) and the width ($\Gamma\in\mathds{R}$) of the state. To some extent this nomenclature is borrowed from the classical physics referring to, e.g., a dumped pendulum which can be pumped if external force is applied at a resonance frequency ($\sim M$) typical to this system, but which can also ``decay'' in a time typical to the system ($\sim 1/\Gamma$). For a historical overview of this analogy see Ref.~\cite{bleckneuhaus2018mechanicalresonance300years}. We note that this crude analogy is agnostic to the quantum nature of fundamental particles and has no further practical meaning as we will see in the next section. The mass and width values of a subset of all known mesons and baryons are shown in Fig.~\ref{fig:MES-BAR-Resonances}. It is seen from this figure that the vast majority of hadrons are indeed unstable, i.e., have non-zero width. Stable hadrons are already mentioned nucleons, pions, kaons and eta mesons. As a side note, we see from this figure a convenient unification of stable and unstable particles as states which only differ in the position of theory parameters in the $M-\Gamma$-plane. This notion will be further extended and formalized in the next section. It is notable that hadrons can be excited/produced in different ways. In collider experiments such as, e.g., Large Hadron Collider~\cite{ATLAS:2008xda} excited hadrons can be produced in the so-called hadronization phase of the final-state interaction. Produced short-lived hadrons (such as pions or kaons) can, however, also then be used as so-called secondary beams hitting a fixed proton or other target see e.g., Klong experiment at Jefferson Laboratory~\cite{KLF:2020gai}. Another popular modern approach is to use photon beams in the so called photo- or electroproduction experiments utilizing either real or virtual photon beams (e.g., ELSA~\cite{Hillert:2006yb} or CLAS~\cite{CLAS:2003umf}). A sample of the photoproduction data encoding several nucleon resonances is depicted in Fig.~\ref{fig:exp-data} adapted form the recent review~\cite{Thiel:2022xtb}. In addition, it is also notable from this figure that hadrons couple differently to different final states. Thus, depending on a specific case, one or another experimental setup may be advantageous.

%%%%%%%%%%%%%%%%%%%%%%%%%%%%%%%%%%%%%%%%%%%%%%%%%%%%%%%%%%%%%%%%
%%%%%%%%%%%%%%%%%%%%%%%%%%%%%%%%%%%%%%%%%%%%%%%%%%%%%%%%%%%%%%%%

%%%%%%%%%%%%%%%%%%%%%%%%%%%%%%%%%%%%%%%%%%%%%%%%%%%%%%%%%%%%%%%%
\section{Resonance parameters}
\label{sec:resonance-parameter}
%%%%%%%%%%%%%%%%%%%%%%%%%%%%%%%%%%%%%%%%%%%%%%%%%%%%%%%%%%%%%%%%

The spectrum of hadrons appears to be very densely populated (Fig.~\ref{fig:MES-BAR-Resonances}) such that different resonances often appear in the same observables at similar energies. Because of the quantum nature of these states, effects of those can further lead to destructive interference and a resonance can indeed appear as a dip in a given observable. Clearly this renders the simple ``bump hunting'' in most cases impossible. \emph{So how then can one disentangle different resonances from each other?} 
The most universal and modern approach to resonances relies on the concept of the so-called $S$-matrix -- originally introduced by W.~Heisenberg~\cite{Heisenberg:1943zz}. In a nutshell, the $S$-matrix is an operator which maps an incoming (remote past) state to an outgoing (remote future) state. Its elements ($S$-matrix element) with respect to the in- and outgoing states with three-momenta $\bm{p}'_1,...,\bm{p}'_n$ and $\bm{p}_1,...,\bm{p}_m$, respectively, are given by 
%%%%%%%%%%%%%%%%
\begin{align}
\langle\bm{p}'_1,...,\bm{p}'_n|S|\bm{p}_1,...,\bm{p}_m\rangle
=:
\langle\bm{p}'_1,...,\bm{p}'_n|(\mathds{1}+\mathrm{i}T)|\bm{p}_1,...,\bm{p}_m\rangle\,,
\end{align}
%%%%%%%%%%%%%%%%
which also defines the $T$-matrix whose elements will be referred to as the scattering amplitude. This is a complex-valued function with respect to the independent kinematical variables, and it encodes everything about the quantum mechanical scattering process. 
%For a general case such kinematical variables can be most conveniently inner products of all involved momenta ($\bm{p}^{(\prime)}_i\bm{p}^{(\prime)}_j$) 
For more formal definitions, see, e.g., Refs.~\cite{Heisenberg:1943zz,gribov_2008,Eden:1966dnq}. Since $S$-matrix elements quantify an overlap of a state before and after the scattering, their squares provide one with a probability of a scattering to happen in the usual sense of quantum mechanics. Thus, they allow to access experimental observables, e.g., total cross sections through $\sigma_{i\to f}\sim|T_{fi}|^2$. At the same time, the $S$-matrix elements indeed can be calculated from a given quantum field theory (QCD, CHPT, ...). Thus, indeed, this approach bridges experiment with theory. Additionally, the $S$-matrix elements obey some mathematical constraints:
\begin{enumerate}
	\item \emph{Crossing symmetry}. In quantum field theories particles can be related to antiparticles. Thus, $S$-matrix element for a $n\to m$ transition can be converted analytically to the one describing $n+1\to m-1$ transitions etc.. That is for example that processes $\pi N\to\pi N$, $\pi\pi\to N\bar N$ are encoded through one and the same function of kinematic variables. For formal equations see, e.g., Ref.~\cite{Schopper:1983hnv}.
	\item \emph{Analyticity}. The aforementioned function is necessarily a meromorphic in the kinematic variables due to causality. For an outline of the reasoning see Ref.~\cite{Mai:2022eur} and for technical details Refs.~\cite{vanKampen:1953rya,Screaton:1969he,Minerbo:1971gg,gribov_2008,Eden:1966dnq}. The important implication of this in view of hadron spectrum research is that:
	\begin{enumerate}
		\item The physical $S$-matrix elements are boundary values to an analytic function in independent kinematic variables, promoted to their complex values, e.g., in complex-valued total energy squared ($s$).
		\item Besides bound states poles at real energies below production thresholds, the complex plane is pole-free.
	\end{enumerate}
	\item \emph{Unitarity}. The $S$-matrix maps in- to the outgoing states. Since the probability to measure anything is per definition unity, completeness and orthonormal property of the physical states yields in the operator language (for more formal derivation see, e.g., Ref~\cite{Eden:1966dnq})
	%%%%%%%%%%%%%%%%
	\begin{align}
		SS^\dagger=\mathds{1} ~\Leftrightarrow~ T-T^\dagger=iTT^\dagger\,.
		\label{eq:S-unitarity}
	\end{align}
	%%%%%%%%%%%%%%%%
	%It is notable that the number of physical states depends on the energy of the system. 
	For practical applications this can be translated to a formal condition on the scattering amplitude by inserting a complete set of intermediate states $|\bm{k}_1,...\bm{k}_N\rangle\langle\bm{k}_1,...\bm{k}_N|$
	%%%%%%%%%%%%%%%%
	\begin{align}
		\label{T-matrix Unitarity}
		\langle \bm{p}'_1,\ldots,\bm{p}'_n|(T-T^\dagger)|& \bm{p}_1,\ldots,\bm{p}_n\rangle=\\
		&
		i\sum_{n=2}^N\int
		\frac{\mathrm{d}^4k_1}{(2\pi)^{4}}\cdots\frac{\mathrm{d}^4k_n}{(2\pi)^{4}}
		\,(2\pi)\delta^+(k_1^2-m^2)\cdots (2\pi)\delta^+(k_n^2-m^2)
		\langle \bm{p}'_1,\ldots,\bm{p}'_n| T^\dagger
		|\bm{k}_1,\ldots\bm{k}_N\rangle\langle\bm{k}_1,\ldots\bm{k}_n|
		%|\{\bm{k}\}\rangle\,,\langle\{\bm{k}\}|
		T| \bm{p}_1,\ldots,\bm{p}_n\rangle \,,\nonumber
	\end{align}
	%%%%%%%%%%%%%%%%
	where $\delta^+(k_i^2-m^2):=\theta(k_i^0)\delta(k_i^2-m^2)$ selecting only positive energy solution. Note that the number ($N$) of individual states in the intermediate states depends on the total energy of the system. That is to say that only states which physically can be realized lead to the imaginary part of the scattering amplitude. Consider for example a generic scattering process of scalar fields (mass $m$). Then, for total energy squared lower than three-body threshold ($4m^2<s<9m^2$) only two-body states ($N=2$) need to be considered. For $9m^2<s<16m^2$ on the other hand, also three-body states need to be included in the intermediate states ($N=3$) etc.. In principle, resolving the integral on the right hand side of the above equation~\eqref{T-matrix Unitarity} is straightforward, when the exact form of the $T$-matrix is known. However, this is rarely the case, instead, the above equation is reformulated to provide a guidance on the construction of the latter. Such as procedure is quite standard for $N=2$ and will be demonstrated below. For higher number of intermediate states, the situation quickly becomes very intricate, see Refs.~\cite{Mai:2017vot,Feng:2024wyg}.
\end{enumerate} 

Taking another look on Fig.~\ref{fig:MES-BAR-Resonances} one realizes that hadronic resonances lie well above the lowest production thresholds. Crossing symmetry constraint discussed above allows one to take data from different channels into account and through this indeed provides an essential constraint on the scattering amplitudes. Especially for the low energy regime this enables one to make high-precision predictions, see, e.g., Ref.~\cite{Ananthanarayan:2000ht} and references therein. Still, so far in most theoretical calculations aiming in extracting meson and baryon resonance parameters crossing symmetry is implemented approximately at most. 

Two other principles (unitarity and analyticity) are unavoidable for hadron spectroscopy, providing for instance a pathway to extract \emph{universal} -- reaction independent resonance parameter. To demonstrate the implications provided through the unitarity condition~\eqref{T-matrix Unitarity}, we proceed with an example of a generic two-body scattering of scalar fields ($\phi(p)\phi(P-p)\to\phi(p')\phi(P-p')$) with equal masses ($m$). The total four-momentum of the system is set to $P=(\sqrt{s},\bm{0})$ corresponding to the center of mass system. Next, one can separate the angular (momentum) and scalar (energy) dependencies off of each other using, e.g.,  orthonormal spherical harmonics $\int d\Omega_{\hat{x}} Y_{lm}(\hat x) Y^*_{l'm'}(\hat x)=\delta_{ll'}\delta_{mm'}$. Here and in the following $\Omega_{\hat x}$ denotes a solid angle with respect to the coordinate $x$. Denoting the matrix element as $T(\bm{p},\bm{p}',s):=\langle \bm{p}',-\bm{p}'|T|\bm{p},-\bm{p}\rangle$, the projection formulas read 
%%%%%%%%%%%%%%%
\begin{align}
& 
	T(\bm{p},\bm{p}',s)=
	4\pi \sum_{lm,l'm'}Y_{l'm'}(\hat p')T_{l'm',lm}(s)Y_{lm}^*(\hat p)\,,
\label{eq:basics:PWdecomp-1}
\\
&
	T_{l'm',lm}(s)=\frac{1}{4\pi}\int d\Omega_{\hat p}\int d\Omega_{\hat p'}
	Y_{l'm'}^*(\hat p')
	T(\bm{p}, \bm{p}', s)
	Y_{lm}(\hat p)\,.
\label{eq:basics:PWdecomp-2}
\end{align}
%%%%%%%%%%%%%%%
Here we have denoted total angular momentum by $l^{(\prime)}=0,\ldots,\infty$ and its third component $m^{(\prime)}=-l^{(\prime)},\ldots,l^{(\prime)}$ with sums running accordingly. Using the latter projection formula~\eqref{eq:basics:PWdecomp-2} on Eq.~\eqref{T-matrix Unitarity} restricting for now to the elastic region ($4m^2<\operatorname{s}<9m^2 \Rightarrow N=2$) one evaluates
%%%%%%%%%%%%%%%
\begin{align}
T_{l'm',lm}(s)-T_{l'm',lm}^*(s)
&=\frac{i}{4\pi}
    \int d\Omega_{\hat p'}d\Omega_{\hat p} 
    Y^*_{l'm'}(\hat p')
    Y_{lm}(\hat p)
    \int \frac{dk^4}{(2\pi)^4}
    T(\bm{p}',\bm{k},s)T^*(\bm{k},\bm{p},s)
    (2\pi)^2\delta^+\left(k^2-m^2\right)\delta^+\left((P-k)^2-m^2\right)
    \\
&= i4\pi
\int \frac{dk_0d{\rm k} {\rm k}^2}{(2\pi)^4}
(2\pi)^2
\delta^+\left(k^2-m^2\right)
\delta^+\left((P-k)^2-m^2\right)
\sum_{l^{(3)}m^{(3)}}T_{l'm',l^{(3)}m^{(3)}}(s)T^*_{l^{(3)}m^{(3)},lm}(s)
\\
&=\frac{i}{\pi}\sum_{l^{(3)}m^{(3)}}T_{l'm',l^{(3)}m^{(3)}}(s)T^*_{l^{(3)}m^{(3)},lm}(s)
\int d{\rm k} \frac{{\rm k}^2}{2E}
\delta^+\left((P_0-E_k)^2-{\rm k}^2-m^2\right)
\\
&=\frac{ip_{\rm cms}}{4\pi\sqrt{s}}
    \sum_{l^{(3)}m^{(3)}}T_{l'm',l^{(3)}m^{(3)}}(s)T^*_{l^{(3)}m^{(3)},lm}(s)\,.
\label{eq:PWunitarity}
\end{align}
%%%%%%%%%%%%%%%
We have used here orthogonality of spheric harmonics, as well as have evaluated $\delta^+$ with respect to $k^0$ and ${\rm k}:=|\bm{k}|$ in order to get to the second line, third and forth line, respectively. We used furthermore $\delta\left(g(x)\right)=\sum_{x_*}\delta(x-x_*)/|g'|_{x=x_*}$ for $g(x_*)=0$, denoting particle energy by $E_k=\sqrt{{\rm k}^2+m^2}$ and the modulus of the three-momentum in the center of mass system as $p_{\rm cms}=\sqrt{s/4-m^2}$. Using now angular momentum conservation ($l=l',\,m=m'$) and Schwarz reflection principle ($T^*(s)=T(s^*)$) one arrives at what is sometimes referred to as the partial-wave unitarity condition
%%%%%%%%%%%%%%%
\begin{align}
&\operatorname{Disc}T_{lm}(s)
	:=\underset{\epsilon\to 0}{\operatorname{lim}} \left(T_{lm}(s+i\epsilon)-T_{lm}(s-i\epsilon)\right)
	=\frac{ip_{\rm cms}}{4\pi\sqrt{s}}|T_{lm}(s)|^2\,,
	\qquad 4m^2<\operatorname{Re}s<9m^2\,,
	\qquad \epsilon>0\,,
\label{eq:basics:2b-PW-unitarity}
\end{align}
%%%%%%%%%%%%%%%
where the limit $\epsilon\to 0$ is evaluated in the last step. This simple derivation can be seen as a baseline for more complex scenarios: (1) The extension to the case of unequal masses is straightforward and is left for the reader as an exercise; (2) For inclusion of non-zero spin fields see, e.g., Refs.~\cite{Jacob:1959at,Schopper:1983hnv}; (3) For higher number of intermediate states, see Ref.~\cite{Eden:1966dnq, Mai:2017vot} and references within. 

To reiterate what we have derived in Eq.~\eqref{eq:basics:2b-PW-unitarity} in words is to say that the unitarity condition for the $S$-matrix Eq.~\eqref{eq:S-unitarity} implies a discontinuity of the scattering amplitude when going from the upper half-plane ($s\in\mathds{C}^+$) to the lower one ($s\in\mathds{C}^-$) and crossing the real axis ($s\in\mathds{R}$) above production threshold. Below it there is no discontinuity\footnote{As discussed before, crossing symmetry relates, e.g., $s$-channel $\pi N\to\pi N$ to $t$-channel $\pi\pi\to N\bar N$ amplitudes. Thus, the discontinuity in the $t$-variable in the latter leads to the so-called left-hand cut in the former. For this introductory discussion we refrain from further discussion of this, but refer the interested reader to a formal discussion in Ref.~\cite{Kennedy:1962ovz}.} implying also that the scattering amplitude is entirely real on the real $s$-axis $T_{lm}(s\in \mathds{R}^{<4m^2})\in \mathds{R}$. This discontinuity or a \emph{branch cut} starting with what is referred to as the \emph{branch point} $s=4m^2$ is called a \emph{unitarity cut}. In general, there are more than one unitarity cuts each one leading to a higher multi-valuedness of the scattering amplitude $T_{lm}(s\in\mathds{C})\in\mathds{C}$. Each of such values defines a Riemann-sheet referred to as first $T_{lm}^{\rm I}(s\in\mathds{C})\in\mathds{C}$,  second $T_{lm}^{\rm II}(s\in\mathds{C})\in\mathds{C}$, etc., building together a continuous Riemann-surface, see Ref.~\cite{Mai:2022eur} for topological interpretation and further details.

Coming back to the resonance parameters, one can show~\cite{Eden:1966dnq} that the usual prescription of the perturbation theory providing a small negative imaginary parts to the masses in propagators corresponds to the upper half-plane ($s\in\mathds{C}^+$). Therefore, the upper-half plane of the first (physical) Riemann sheet is the one connected smoothly to the physical scattering amplitude which allows one to calculate measurable quantities for real energies. As discussed above, poles in $T_{lm}$ on this Riemann sheet can only appear on the real axis below production threshold, corresponding to the bound states. On the unphysical Riemann sheets no such constraint exists and poles are allowed there. One refers to a \emph{resonance pole} at $s_*\in \mathds{C}$ for $\operatorname{Im} s_*\ne 0$ which always appear as a conjugate pairs. Since the second Riemann sheet is connected smoothly to the first one along the positive energy-axis above threshold, no pole is allowed on the real energy-axis. However, below threshold this is again allowed, the corresponding state is referred to as a \emph{virtual bound state}. 

To summarize, resonance parameters are encoded in the analytical properties of the $S$-matrix which we can also associate to the old fashioned mass and width of an excited state. The physical coupling ($g$) to an individual final state (renormalized and in general complex-valued quantity) is recovered from the residuum of the scattering amplitude at the pole position $s_*$ as 
%%%%%%%%%%%%%%%%%
\begin{align}
	\sqrt{s_*}=M-i\Gamma/2\,, 
	\qquad
	g^2=\underset{s\to s_*}{\lim}(s-s_*)T_{lm}(s)\,.
\end{align}
%%%%%%%%%%%%%%%%%
This identification allows one to identify the lifetime of a decaying states with its width as $\tau=1/\Gamma$ see Ref.~\cite{Willenbrock:2022smq}. The key point is that this procedure unites the concept of bound states, virtual bound states and resonances. In that, the obtained pole positions do not depend on the particular choice of the reaction through which an excited hadron might have been created (scattering, photo-production experiment, etc.) as discussed in the previous section Sec.~\ref{sec:theory-exp}.  As a matter of fact, we now have tools to tune parameters of QCD observing pole movement with respect to these parameters on the Riemann surface and in some cases, indeed, witnessing transmutation between different state types, see next section.

%%%%%%%%%%%%%%%%%%%%%%%%%%%%%%%%%%%%%%%%%%%%%%%%%%%%%%%%%%%%%%%%
\section{Applications/Examples}
\label{sec:applications}
%%%%%%%%%%%%%%%%%%%%%%%%%%%%%%%%%%%%%%%%%%%%%%%%%%%%%%%%%%%%%%%%

% In the following we review a few approaches used frequently in the hadron spectroscopy to access resonance parameter.

\subsection{Amplitude types}

Historically, the primer tool in accessing resonance parameters from experiments has been the so-called Breit-Wigner parametrization~\cite{Wigner:1946zz}. Strictly speaking this parametrization is applicable to only a very small number of cases, that is for narrow resonances located far from production thresholds. Specifically, the total cross section $\sigma$ for the scattering of a plane wave (wave vector $\bm{k}\in\mathds{R}^3,~{\rm k}:=|\bm{k}|$) with an angular momentum $l$ is parametrized as
%%%%%%%%%%%%%%%%
\begin{align}
\label{eq:BW}
	\sigma(s)\propto\frac{4\pi}{{\rm k}^2}(2l+1)\frac{\Gamma^2/4}{(\sqrt{s}-M)^2+\Gamma^2/4}\,.
	%\quad\text{and}\quad
	%\tan\delta\propto -\frac{\Gamma_\ell}{2(E-E_R)}\,.
\end{align}
%%%%%%%%%%%%%%%%
This explains again the name ``width'' for $\Gamma$ representing twice the distance from the peak (at $M$) to the half of the maximal value of the total cross section. In a slightly modified version of this -- the so-called Flatt\'e{} parametrization~\cite{Flatte:1976xu} one can also take into account the coupled-channel effects. Consider a state $\Psi$ which can decay to $\phi_1+\phi_2$ (mass $m_{1/2}$) or to some heavier states $\Phi_1+\Phi_2$ (mass $M_{1/2}$) such that $|M_1+M_2|>|m_1+m_2|$ and $|M_1-M_2|<|m_1+m_2|$. Then the cross section above the lowest threshold ($s>(m_1+m_2)^2$) is parametrized as
%%%%%%%%%%%%%%%%
\begin{align}
	\label{eq:Flate}
	\sigma(s)
	\propto
	\left|
		\frac{m_R\sqrt{\Gamma_{\phi_1\phi_2}\Gamma_{\Phi_1\Phi_2}}}	{m_R^2-s-im_R(\Gamma_{\phi_1\phi_2}+\Gamma_{\Phi_1\Phi_2})}
	\right|^2
	&
	\qquad
	\Gamma_{\Phi_1\Phi_2}=
	\left\{
    \begin{array}{lr}
        g_{\Psi\to \Phi_1\Phi_2}i\sqrt{\frac{(-s+(M_1+M_2)^2)(s-(M_1-M_2)^2)}{4s}}, & \text{if } 	s<(M_1+M_2)^2\\
		g_{\Psi\to \Phi_1\Phi_2}\sqrt{\frac{(s-(M_1+M_2)^2)(s-(M_1-M_2)^2)}{4s}},  & \text{if } 	s>(M_1+M_2)^2
    \end{array}
	\right.\\
	&
	\qquad
	\Gamma_{\phi_1\phi_2}=
		g_{\Psi\to \phi_1\phi_2}\sqrt{\frac{(s-(m_1+m_2)^2)(s-(m_1-m_2)^2)}{4s}}\,.
\end{align}
%%%%%%%%%%%%%%%%
We have denoted here the dimensionless coupling constant to individual channels by $g$. A typical application example is the $a_0(980)\to\{\pi\eta,K\bar K\}$, see Refs.~\cite{Kerbikov:2004uz,Baru:2004xg} and for related discussions Refs.~\cite{Dong:2020hxe,Sone:2024nfj}. Note the appearance of the $\sqrt{(s-(m_1+m_2)^2)(s-(m_1-m_2)^2)/4s}$ function which is identical to the modulus of the three-momentum in the center of mass system for the case of unequal masses. Thus, also two-body unitarity is explicitly build in this form as it will become clear below, see Eq.~\eqref{eq:basics:2b-PW-unitarity} and Eq.~\eqref{eq:PW-unitarity-Kmatrix}. 

When only two-particle states are involved in the scattering process the unitarity condition Eq.~\eqref{eq:basics:2b-PW-unitarity} directly provides a guidance on the form of the partial-wave amplitudes. Specifically,
%%%%%%%%%%%%%%%%
\begin{align}
	\label{eq:PW-unitarity-Kmatrix}
	\operatorname{Im} T_{l}^{-1}(s)=-\frac{p_{\rm cms}}{8\pi\sqrt{s}}
	\qquad
	\Longrightarrow
	\qquad
	T_{l}(s)=\frac{8\pi\sqrt{s}}{K^{-1}_{l}(s)-ip_{\rm cms}}\,.
\end{align}
%%%%%%%%%%%%%%%%
Clearly, the complexity in defining the scattering amplitude has been simply relegated here to defining an unknown real-valued quantity ($K_l$) the so-called $K$-matrix. It is interesting to note that, while we demand that $T_{l}(s\in \mathds{C})$ is meromorphic, $K_{l}(s)$ is in general not (unless it is constant). Still, one can now anticipate how multi-channel scenario can be implemented by simply extending the above quantities to matrices of an appropriate size. In the simplest case, some algebraic parametrization (e.g.,~Pad\'e, Chew-Mandelstam forms, see Ref.~\cite{Baker_Graves-Morris_1996}) of the $K$-matrix can be assumed fitting its parameters to the available data. If enough data is available, analytic extrapolation to the second Riemann-sheet become viable and resonance parameters can be extracted through pole positions as described above. In the past, such a workflow, indeed, led for example to the discovery~\cite{Dalitz:1967fp} of the famous $\Lambda(1405)$-resonance below the production threshold, for history and details see the dedicated review~\cite{Mai:2020ltx}. Another important advantage is that the $K$-matrix can be related to phase-shift ($\delta$) as 
\begin{align}
	&1+iT_l(s)=e^{2i\delta_l(s)}=S_l(s)
	% &1-2\sin^2(\delta_l(s))+i2\cos(\delta_l(s))\sin(\delta_l(s))=1-Im T_l(s)+iReT_l(s)\\
	% &2\sin^2=ImT ~~~\&~~~ 2\cos\sin=Re T\\
	~~\Longrightarrow~~
	\cot(\delta_l(s))=\frac{\operatorname{Re} T_l(s)}{\operatorname{Im}T_l(s)}
	~~\Longleftrightarrow~~
	K^{-1}_l(s)=p_{\rm cms}\cot(\delta_l(s))\,,
	% \\
	% &
	% \overset{ERE}{\longleftrightarrow}
	% p_{\rm cms}\cot\delta=-\frac{1}{a}+r\frac{p_{\rm cms}^2}{2}+\ldots\,.
	\label{eq:K-matrix-pcotdelta}
\end{align}
related through the effective range expansion $p_{\rm cms}\cot(\delta)=-\frac{1}{a}+r\frac{p_{\rm cms}^2}{2}+\ldots$ to  scattering length ($a$) and effective range ($r$). Note that while normalization of the $T$-matrix differs in the literature, it cancels exactly in the second equation. It is worth noting that the use of the $K$-matrix form is of large relevance for lattice hadron spectroscopy programs dealing with finite-volume effects. Qualitatively, this is simply realized because unitarity accounts of all on-shell two-body states, separating them off of the short-range and off-shell contributions in the $K$-matrix. It is only the former which produce non-negligible (polynomial vs. exponential suppression with respect to the finite box size $L$) finite-volume effects. Thus, the latter can be extracted either from finite-volume lattice spectra and compared to the one fixed through experimental results.
% For more details and explicit formulas see Chapter~[{\color{red}20019}].

A large class of amplitudes is produced through unitarization, including approaches such as Chiral Unitary Approaches (UCHPT), $N/D$ or Inverse Amplitude Methods. While there are various ways to introduce this methodology, we stay here with a diagrammatic language and start for pedagogical reasons from the scalar $\phi^4$-theory. The Lagrangian of this theory reads
%%%%%%%%%%%%%%%%
\begin{align}
	\label{eq:phi4-Lagrangian}
	\mathcal{L}=\frac{1}{2}\left(\partial^\mu\phi \partial_\mu\phi-M^2\phi^2\right)
	-\frac{\lambda}{4!}\phi^4\,,
\end{align}
%%%%%%%%%%%%%%%%
with $M$ the particle mass and $\lambda$ a coupling constant. This simple form of the interaction leads to the fact that the scattering amplitude for the process $\phi(p_1)\phi(p_2)\to \phi(p_1')\phi(p_2')$ separates into an infinite series of Feynman diagrams ordered in powers of $\lambda$. Note that in Chiral Perturbation Theory one would order this series by the corresponding chiral order. Ignoring the mass-renormalization due to self-energy diagrams, symmetry factors etc. for simplicity, the first terms of this series read
%%%%%%%%%%%%%%%%
\begin{align}
	T_1(\bm{p}_1',\bm{p}_1,P)&=-\lambda\,,\nonumber\\
	T_2(\bm{p}_1',\bm{p}_1,P)&=-\lambda^2\left(\tilde G(p_1+p_2)+\tilde G(p_1-p_1')+\tilde G(p_1-p_2')\right)\,,
	~~~~~\text{with}~~~~~	
	\tilde G(p)=\int\frac{d^dk}{(2\pi)^d}\frac{i}{k^2-M^2+i\epsilon}\frac{1}{(k-p)^2-M^2+i\epsilon}\,,
	\label{eq:phi4-expansion}
	\\
	% T_3(\bm{p}_1',\bm{p}_1,P)&=-\lambda^3\left(\ldots\right)\,,\nonumber
	T_3(\bm{p}_1',\bm{p}_1,P)&=-\lambda^3\left(\tilde G(p_1+p_2)^2+\ldots\right)\,,\nonumber
\end{align}
%%%%%%%%%%%%%%%%
denoting again the total four-momentum by $P=p_1+p_2=p_1'+p_2'$ and the one-loop function in $d$ dimensions by $\tilde G(p)$. The latter is logarithmically divergent for $d=4$, but can be treated in the usual sense of perturbative renormalization, i.e., absorbing them in the renormalized particle mass $M$, the coupling $\lambda$ and the field renormalization factor $Z$ at any given order. In dimensional regularization and the $\overline{\rm MS}$ subtraction scheme the finite part of the loop integral becomes
%%%%%%%%%%%%%%%%
\begin{align}
	\tilde G(x)=\frac{1}{16\pi^2}\left(-1+2\ln\left( \frac{M}{\mu}\right)-\frac{4q(x)}
	{\sqrt{x^2}}{\rm tanh}^{-1}\left(\frac{2q(x)\sqrt{x^2}}{4M^2-x^2}\right)\right)
	\quad\text{with}\quad
	q(x)=\sqrt{x^2/4-M^2}
	\,,
\label{eq:dimregloop}
\end{align}
%%%%%%%%%%%%%%%%
where $\mu$ denotes the regularization scale. One notes that the expression for general values of $x^2\in\mathds{R}$ the above expression is complex-valued with the imaginary part existing only for $x^2>4M^2$. Thus, in the energy region of interest $P^2=s>4M^2$ (physical region) only $\tilde G(p_1+p_2)=:\tilde G(s)$ can lead to the imaginary part of $T_2$ in Eq.~\eqref{eq:phi4-expansion} reading
%%%%%%%%%%%%%%%%
\begin{align}
	\operatorname{Im}T_2=-|T_1|^2 \operatorname{Im}\tilde G(s)=|T_1|^2 \frac{q(s)}{8\pi\sqrt{s}}\,.
\end{align}
%%%%%%%%%%%%%%%%
We have used first leading order result from Eq.~\eqref{eq:phi4-expansion} in the last step. The arising mixing of orders of expansion in this relation is sometimes referred to as perturbative unitarity, to be understood in contrast with the full unitarity condition~\eqref{eq:PW-unitarity-Kmatrix}. One can restore the latter using for example the \emph{Bethe-Salpeter equation} (BSE) 
%%%%%%%%%%%%%%%%
\begin{align}
	T(p_1',p_1;p)=V(p_1',p_1;p)-\int\frac{d^dk}{(2\pi)^d}V(p_1',k;p)G(k;p)T(k,p_1;p)\,,
	~~\quad~~
	G(k,p):= \frac{i}{k^2-M^2+i\epsilon}\frac{1}{(p-k)^2-M^2+i\epsilon}\,,
	\label{eq:BSE}
\end{align}
%%%%%%%%%%%%%%%%
which effectively corresponds to a re-summation of a certain interaction kernel $V$. Most commonly one restrict the latter to contact interaction (e.g., $V=T_1$) leading to an infinite series of Feynman diagrams exactly corresponding to the first terms on the right hand side of the perturbative expansion in Eq.~\eqref{eq:phi4-expansion}. There are some examples where the solution of the full $d$-dimensional integral equation has been obtained~\cite{Nozawa:1989pu, Lee:1991pp, vanAntwerpen:1994vh, Borasoy:2005zg, Bruns:2010sv, Ruic:2011wf, Mai:2013cka} but in general such a solution does not exist. One simplification -- often referred to as the \emph{quasipotential} or
\emph{Gross} equation, see, e.g., Ref.~\cite{Gross:1969rv} -- is achieved by assuming that neither $T$ nor the interaction kernel $V$ have singularities in $k^0$ on the real axis, allowing to perform the $k^0$ integral using Cauchy's theorem. This effectively puts the intermediate particles on their mass-shell $k^2=M^2$ reducing dimensionality of the integral equation but destroying manifest covariance. Next, projecting to partial waves one obtains even simpler, algebraic equation
%%%%%%%%%%%%%%%%
\begin{align}
	T_l(s)
	=\tilde V_l(s)-\tilde V_l \tilde G(s) T_l(s)
	=\frac{1}{1+\tilde V_\ell(s)\tilde G(s)}\tilde V_\ell(s)\,.
\label{eq:unitarized-amp}
\end{align}
%%%%%%%%%%%%%%%%
referred most-commonly to as the \emph{unitarized scattering amplitude}. Since it fulfills unitarity exactly it also can be related to the previously obtained $K$-matrix form via $K_l^{-1}(s)=\tilde V_l^{-1}(s)+{\rm Re}~\tilde G(s)$.
Because not all topologies from the perturbative expansion in Eq.~\eqref{eq:phi4-expansion} can be reproduced in this way, the renormalization procedure cannot be carried out in the usual sense. Related residual regularization scale dependence but also truncation of the potential $V$ inevitably induces some model-dependence in the unitarized approaches. Still, this methodology provides a path from, e.g., Chiral Lagrangian (incorporating for instance symmetries of QCD) to unitary scattering amplitudes. These can be compared with experimental or Lattice QCD results, and provide ultimately (through analytic continuation to the unphysical Riemann sheets) access to universal parameters of resonances. Examples of the light excited hadrons successfully studied and in some cases even discovered~\cite{Oller:2000fj} through such approaches includes: $\rho(770)$, $f_0(500)$, $N(1335)$, $N(1650)$, $\Lambda(1405)$ or $\Lambda(1380)$. For a discussion of various resonance schemes in collider experiments see, e.g., Ref.~\cite{Ge:2024bjp}. For reviews on heavier part of the spectrum see Refs.~\cite{Mai:2022eur,Guo:2017jvc,Chen:2022asf,Meng:2022ozq}.

%%%%%%%%%%%%%%%%%%%%%%%%%%%%%%%%%%%%%%%%%%
\subsection{Resonance properties}
\label{subsec:resonance_properties}
%%%%%%%%%%%%%%%%%%%%%%%%%%%%%%%%%%%%%%%%%%

After the general framework for finding reaction independent parameters of resonances has been outlined, we may step back and reevaluate the goals of hadron spectroscopy. One may ask for example, if \emph{the only thing left is to perform more experiments, deducing more intricate and complete parametrizations of transition amplitudes filling up the table of resonance parameters?} Questions which go beyond this endeavor usually refer to the \emph{``nature''} of resonances, which in a sense is similar to what motivated Bohr's picture in the atomic theory, see Sec.~\ref{sec:intro}. Clearly, attempting answering this question one can quickly enter the realm of philosophy. However, we showcase in the following several cases a path to tell something systematic observation based about the inner structure of resonances. These more advanced topics serve as a first glimpse into the matter, while formal details can be found in the provided references. 

First and foremost question about the nature of a given state is whether the state is composite or elementary. It seems that in general this question cannot be decided in a model independent way. However, when certain conditions are fulfilled the probability of finding a given state in an elementary one can, indeed, be related to observables. The most famous example of this is the deuteron, a bound system of a proton and a neutron, discussed in Ref.~\cite{Weinberg:1965zz}. There, the probability was related to the low-energy scattering parameter -- scattering length ($a$) and effective range ($r$) --  accessible through low-energy experiments. Specifically, neglecting finite-range interactions it was derived that 
%%%%%%%%%%%%%%
\begin{align}
	a=2\frac{1-Z}{2-Z}R
	\text{~~and~~}
	r=\frac{Z}{Z-1}R\,.
	% \quad
	% \overset{ERE}{\longleftrightarrow}
	% \quad
	% p_{\rm cms}\cot\delta=-\frac{1}{a}+r\frac{p_{\rm cms}^2}{2}+\ldots\,,
\end{align}
%c.f., Eq.~\eqref{eq:K-matrix-pcotdelta} in the so-called effective range expansion (ERE). 
The deuteron radius is related to the nucleon mass ($m$) and deuteron binding energy ($E_B$), which are known ($R=(mE_B)^{-1/2}\approx4{\rm fm}$), while the deuteron wave function renormalization constant ($Z$) vanishes for purely composite particles~\cite{Houard,Vaughn:1961poz}. In case of the deuteron, the low-energy data necessitates a very small value of $Z$ thus the probability to find it in the elementary state is very low, too. The defined procedure can be applied to other systems as long as the particle is stable or at least very narrow and it couples to two-body channels in the $S$-wave. For more details 
%see Chapter~{\color{red}20025}, for 
and generalizations of this procedure see Refs.~\cite{Li:2021cue,Sekihara:2014kya,Hyodo:2011qc} and for related reviews Refs.~\cite{Hyodo:2013nka,Oller:2017alp,Guo:2017jvc,Kinugawa:2024crb}.

\begin{figure}[t]
	\centering
	\includegraphics[width=0.55\linewidth]{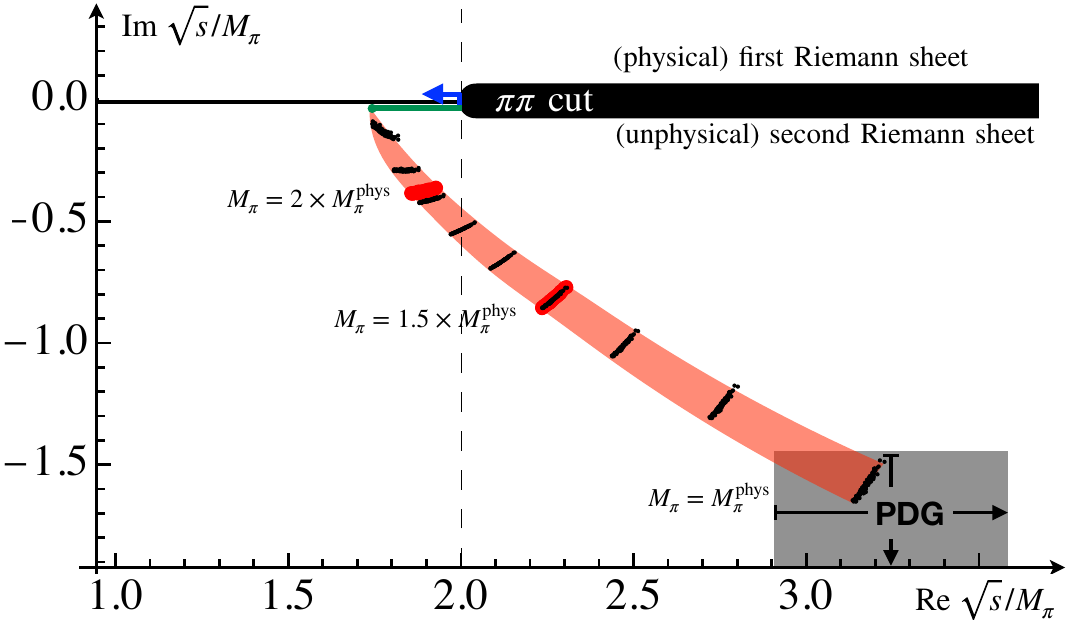}
	\includegraphics[width=0.4\linewidth]{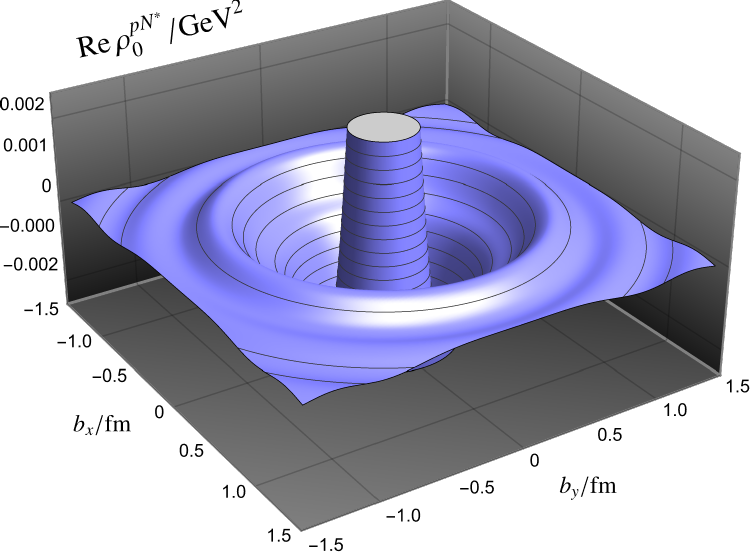}
	\caption{
		Left panel: Pole trajectory of the $f_0(500)$-meson as a function of the employed pion mass. Parameters of the model have been determined at two pion masses $M_\pi\approx 224, 315\,{\rm MeV}$, corresponding pole positions at these pion masses are depicted by thick red dots. The spread of these represents statistical uncertainty of the procedure. Extrapolation to lighter/heavier pion masses is depicted by black dots and colored bands. With increasing pion mass $f_0(500)$ becomes lighter and narrower resonance, becoming a virtual bound state (green curve) at $M_\pi\approx 2.5\times M_\pi^{\rm phys}$. At $M_\pi\approx 3\times M_\pi^{\rm phys}$ the pole moves to the first Riemann sheet and becomes a bound state (blue curve), its binding energy grows further with the increasing pion mass.\\
		Right panel: Quark charge density inducing the transition $N\to N(1440)$. The typical zero transition is a typical sign for a radial excitation of three quarks from the ground state. For more details see the original publication~\cite{Wang:2024byt}. 
		}
	\label{fig:f0500-TFF}
\end{figure}

On a more genral basis, consider two models for an emergence of a specific resonance phenomena. Let's assume that both models are equally successfully describing the available experimental data, but rely on different degrees of freedom generating the resonant state (e.g., quarks vs. hadron-hadron interactions).  It seems now reasonable to regard the one model as more realistic which withstands verification in (loosely speaking) larger generalized parameter-space of dimension-ful quantities such as energies or masses. Obviously, the masses of hadrons and quarks are fixed in our universe, but can be varied in Lattice QCD which has various technical and computational reasons discussed in, e.g., Ref.~\cite{Mai:2021lwb}. The actual example of an outcome of such an exploration through a variant of the previously described UCHPT approach is shown in Fig.~\ref{fig:f0500-TFF}. There, the parameters of the model for the $f_0(500)$-meson are determined through fits to the Lattice QCD results~\cite{Guo:2018zss,Mai:2019pqr} at two unphysical pion masses ($M_\pi^2\propto m_q$, see e.g.~\cite{Gasser:1983yg}). Then, the latter is varied to higher/lower values including also the physical point ($M_\pi\approx 139~{\rm MeV}$). One notes that in the the latter case a post-diction of the $f_0(500)$ resonance parameter at the physical point, indeed, agrees nicely with the PDG~\cite{ParticleDataGroup:2024cfk}. Going to heavier pion masses, one notes further that at around $M_\pi\approx 2.5\times M_\pi^{\rm phys}$ the resonance pole position becomes entirely real staying, however, on the second Riemann sheet -- the state becomes a virtual bound state. Increasing the pion mass further, the pole moves closer to the two-particle threshold from below. It hits then the branch point $\sqrt{s}=2M_\pi$ and proceeds on the first Riemann sheet as a bound state pole with increasing binding energy. In conclusion, this scenario underlines again the advantage of the unified picture of resonances, bound states and virtual bound states, altogether represented through the analytic structures of the transition amplitudes. Meanwhile, similar Lattice QCD backed studies have also been carried out for baryon resonances~\cite{BaryonScatteringBaSc:2023zvt,Guo:2023wes,Zhuang:2024udv} and three-body resonances such as $\omega(782)$-meson~\cite{Yan:2024gwp}.

%%%%%%%%%%%%%%%%%%%%%%%%%%%%%%%%%%%%%%%%%%%%%%%%%%%%%%%%%%%%%%%%
%%%%%%%%%%%%%%%%%%%%%%%%%%%%%%%%%%%%%%%%%%%%%%%%%%%%%%%%%%%%%%%%

Finally, there is also an interesting avenue of probing the structure of the excited states through the modern meson electroproduction experiments at, e.g., the Jefferson Laboratory~\cite{CLAS:2002zlc,CLAS:2022yzd,CLAS:2008agj}. There, and similarly to the discussed photoproduction experiments, a nucleon is excited through photon beams decaying to some asymptotically stable final states. Still, going beyond this, the photon used there is virtual, i.e., it has non-zero  virtuality $Q^2:=q_0^2-\bm{q}^2\ne0$. In the sense of the previous paragraph, this new variable provides another independent dimension to study resonance properties. Corresponding fundamental quantities of such a process are the so-called Helicity amplitudes~\cite{Berends:1967vi} ${\cal H}(s,Q^2)$ whose residua at the resonance poles $s\to s_{\rm pole}$ allow one to access the so-called transition form factors $H(Q^2)$. The $Q^2$ dependence of the latter is determined by the experimental data and can be interpreted through, e.g., functional methods described in Sec.~\ref{sec:theory-exp} and Refs.~\cite{Eichmann:2016yit,Ramalho:2023hqd}. In some specific reference frame (light front) these form factors can further be used to define the quark charge density $\rho_0^{pN^*}$ inducing the transition from the ground state nucleon ($p$) to the excited one ($N^*$), see Refs.,~\cite{Tiator:2009mt,Ramalho:2023hqd}. An example of the latter for the first excited state of the nucleon is shown in the right panel of the Fig.~\ref{fig:f0500-TFF} being determined through the analysis of the current electroproduction data~\cite{Wang:2024byt}. Here again we see how the notion of resonance poles is leading to deeper insights into the structure of excited states.

%\clearpage
%%%%%%%%%%%%%%%%%%%%%%%%%%%%%%%%%%%%%%%%%
\section{Conclusions}
\label{sec:conclusions}
%%%%%%%%%%%%%%%%%%%%%%%%%%%%%%%%%%%%%%%%%

Understanding the spectrum of excited states can be seen as an ultimate test for our understanding of fundamental interactions of nature. This was equally true for the atomic theory establishing quantum mechanics as unavoidable concept and holds for the still ongoing research on spectroscopy of hadrons. The exploration of the latter led to a tremendous progress of our understanding of strong interaction through theoretical and experimental research. In the present article, we have reviewed current state of the art of the latter and provided a brief review of the modern concepts. Specifically, we have focused on techniques rooted in the $S$-matrix theory allowing one to define reaction-independent resonance parameter. Several hands-on tools have been introduced not only as a glossary for future research but also to motivate advanced reading in the provided references. 

With the new already commissioned facilities around the globe as well as the development and refinement of the theoretical techniques, the hadron spectroscopy remains an active research area. Future questions include frog's perspective questions such as quantification of the gluonic degrees of freedom or bird's view questions such as deducing the minimal content of the hadron spectrum and its patterns.

\begin{ack}[Acknowledgments]%
	We thank Feng-Kun Guo and Peter Bruns for useful discussions and comments to the manuscript. 
	The work on this manuscript was funded through the Heisenberg Programme by the Deutsche Forschungsgemeinschaft (DFG, German Research Foundation) - 532635001. Some material is based upon work supported by the National Science Foundation under Grant No. PHY-2310036 and the U.S. Department of Energy, Office of Science, Office of Nuclear Physics under Award Number DE-SC0016582.
\end{ack}

%%%%%%%%%%%%%%%%%%%%%%%%%%%%%%%%%%%%%%%%%%%%
%% Optional: A list of references to other relevant works/articles/websites which are not cited in the text but that would further enhance a readers understanding of this topic
% \seealso{article title article title}

%%%%%%%%%%%%%%%%%%%%%%%%%%%%%%%%%%%%%%%%%
%% Mandatory: Bibliography using bibtex 
\bibliographystyle{Numbered-Style} %% for Numbered Reference Style
\bibliography{BIB, NON-INSPIRE}

\end{document}